\documentclass[
reprint,
superscriptaddress,
showkeys,
amsmath,amssymb,
aps,
prapplied,
]{revtex4-2}

\usepackage{graphicx}
\usepackage{dcolumn}
\usepackage{bm}

\usepackage{lipsum}
\usepackage{xcolor}

\begin{document}

\preprint{APS/123-QED}

\title{Pressure Dependence of Ferromagnetism in Uranium Hydride UH$_3$ up to 20~GPa Revealed by NV Centers Magnetometry}

\author{Valentin Schmidt}
\author{Loïc Toraille}
\author{Florent Occelli}

\affiliation{CEA DAM, DIF, F-91297 Arpajon, France}
\affiliation{Université Paris-Saclay, CEA, Laboratoire Matière en Conditions Extrêmes, 91680 Bruyères-le-Châtel, France}

\author{Jean-François Roch}

\affiliation{Université Paris-Saclay, CNRS, ENS Paris-Saclay, Centralesupelec, LuMIn, 91190 Gif-sur-Yvette, France}

\author{Paul Loubeyre}

\affiliation{CEA DAM, DIF, F-91297 Arpajon, France}
\affiliation{Université Paris-Saclay, CEA, Laboratoire Matière en Conditions Extrêmes, 91680 Bruyères-le-Châtel, France}


\begin{abstract}

The origin of ferromagnetism in the stable $\beta\text{-UH}_3$ phase is still debated. With a high Curie temperature and a short uranium-uranium interatomic distance, $\beta\text{-UH}_3$ is placed outside the known systematics of uranium compounds. Pressure provides a powerful means of tuning magnetism toward its suppression, thereby offering valuable insights into its underlying mechanisms, yet the high-pressure behavior of $\beta\text{-UH}_3$ has remained largely unexplored. Here, we combine in the Diamond Anvil Cell (DAC) the development of pure $\beta\text{-UH}_3$ synthesis and widefield nitrogen-vacancy (NV) center magnetometry to measure the pressure dependence of the Curie temperature in $\beta\text{-UH}_3$ up to about 20~GPa. We present two data analysis methods, vector magnetic field reconstruction and statistical analysis of the optically detected magnetic resonance (ODMR) response, which enable us to directly image the magnetic dipole of $\beta\text{-UH}_3$ during isobaric warming. We observe a linear decrease in the Curie temperature with pressure, yielding a slope of $\text{d}T_C/\text{d}P=-3.86\,(8)$~K/GPa. Ferromagnetism in $\beta\text{-UH}_3$ is predicted to reach 0~K at approximately 44~GPa, where a quantum critical point may emerge. 

\end{abstract}


\maketitle


\section{Introduction}

$\beta\text{-UH}_3$ was the first reported 5f-electron ferromagnet~\cite{gruen_magnetic_1955}. A large saturation magnetization of about 1 Bohr magneton per uranium atom and a high Curie temperature of $T_C \sim 175$~K were observed~\cite{gruen_magnetic_1955,rundle_structure_1947}. $\beta\text{-UH}_3$ crystallizes in a cubic structure with a uranium sublattice of the A15 type (also known as $\beta$-W), while the hydrogen atoms occupy the tetrahedral interstitial sites~\cite{rundle_structure_1947,rundle_hydrogen_1951}. A second metastable phase, $\alpha\text{-UH}_3$, was subsequently discovered~\cite{mulford_new_1954}. It exhibits similar ferromagnetic properties~\cite{tkach_electronic_2015,kyvala_electrons_2022,devanaboina_stabilization_2026} but crystallizes in a different cubic structure of the $\text{Cr}_3\text{Si}$ type.

The magnetic behavior of actinide compounds has long been considered to be governed by the shortest actinide–actinide distance. This empirical correlation, introduced by Hill, is commonly expressed through the so-called Hill limit~\cite{hill1970,Sechovsky1998}: below a critical distance, strong direct overlap between neighboring 5f orbitals favors itinerant, Bloch-like 5f electronic states and suppresses local-moment magnetism, whereas larger distances favor 5f-electron localization and magnetic ordering. For uranium compounds, this critical U–U distance is approximately 350~pm~\cite{boring_plutonium_2000}. The shortest U–U distance in $\alpha\text{-UH}_3$ is about 360~pm~\cite{troc_discovery_1995}, slightly above the Hill limit and therefore consistent with its ferromagnetic behavior. By contrast, $\beta\text{-UH}_3$ exhibits strong ferromagnetism despite a shorter U–U distance of about 330~pm. It therefore represents a striking exception to the Hill criterion, demonstrating that the U–U separation alone is insufficient to account for magnetic ordering in uranium hydrides.

The apparent violation of the Hill criterion in $\beta\text{-UH}_3$ is generally attributed to the modification of the uranium electronic structure induced by hydrogen. Density functional theory calculations~\cite{kyvala_electrons_2022,havela_hydrogen_2023} and experimental studies on $\beta\text{-UH}_3$~\cite{grunzweig-genossar_nuclear_1970,barash_nmr_1984,wilhelm_magnetism_2018,tereshina-chitrova_synthesis_2023,koloskova_5_2024} and $\beta\text{-UD}_3$~\cite{lawson_vibrational_1990,lawson_magnetic_1991} indicate that hybridization with H-1s states induces a substantial charge transfer from uranium to hydrogen. This charge redistribution depopulates the uranium 6d states and consequently weakens the hybridization between the uranium 5f and 6d manifolds. The resulting narrowing of the 5f band~\cite{kyvala_electrons_2022} favors the localization of the 5f electrons and the formation of magnetic moments, despite the short U–U distance. Within this picture, the relevant parameter is no longer the geometric U–U distance alone but rather the effective overlap between neighboring uranium 5f states, which depends on both the U–U separation and the spatial extent of the 5f orbitals. The hydrogen-induced redistribution of the uranium electronic states reduces this effective overlap, allowing localized magnetic moments to persist below the conventional Hill limit~\cite{koloskova_hydrogen_2021}. A quantitative theoretical description of the magnetic properties of $\beta\text{-UH}_3$ nevertheless remains challenging since the localized and itinerant 5f configurations are separated by very small total-energy differences. Consequently, the calculated magnetic ground state is highly sensitive to the theoretical treatment of electronic correlations.

Chemical substitution has been used to probe the sensitivity of $\beta\text{-UH}_3$ magnetism to changes in the uranium electronic environment and in charge transfer across the U–H bond. Studies of $\text{-UH}_3$-based alloys containing various transition-metal elements~\cite{tkach_electronic_2015,devanaboina_stabilization_2026,koloskova_hydrogen_2021,tkach_ferromagnetism_2013,havela_uh3-based_2016,havela_strong_2016,havela_xps_2020,kim-ngan_superconductivity_2018,buturlim_laves_2018} have shown that the ferromagnetic state remains robust, with the ordered magnetic moment and Curie temperature generally only moderately affected by substitution. In some cases, however, the magnetic order is further stabilized, with $T_C$ reaching approximately 200~K in Mo-containing compounds~\cite{havela_uh3-based_2016}. This enhancement is consistent with a role for charge redistribution and orbital hybridization in stabilizing the ferromagnetic state. Nevertheless, chemical alloying changes the composition and introduces local structural and chemical disorder, making it difficult to isolate the underlying microscopic mechanism. Hydrostatic pressure, by contrast, continuously tunes the U–U spacing, the U–H bond lengths, and the degree of orbital hybridization without introducing compositional disorder. It therefore provides a more direct probe of the microscopic origin of ferromagnetism in $\beta\text{-UH}_3$ and, more broadly, of 5f-electron magnetism in actinide materials.

High-pressure experiments on $\beta\text{-UH}_3$ are particularly challenging because of the compound's extreme air sensitivity and pyrophoric nature~\cite{le_guyadec_pyrophoric_2010}, which requires all sample preparation and handling to be performed in an inert-atmosphere glovebox. In addition, the small sample volume accessible in a diamond anvil cell (DAC) at pressures of several gigapascals necessitates highly sensitive magnetic probes to detect the weak magnetic signal. To date, only one high-pressure magnetic study has been reported for pure $\beta\text{-UH}_3$~\cite{andreev_magnetic_1998}. Measurements up to 0.8~GPa revealed a rapid quadratic decrease of $T_C$ with pressure, which was interpreted as a signature of itinerant 5f magnetism. The pressure dependence of $T_C$ was also investigated in the Mo-alloyed hydride $(\text{UH}_3)_{0.82}\text{Mo}_{0.18}$ up to 3.2~GPa~\cite{prchal_pressure_2020}. In that case, $T_C$ decreased linearly with pressure, but with a smaller slope than expected for a typical 5f-band ferromagnet, suggesting a more localized character of the 5f electrons. High-pressure X-ray diffraction measurements have further shown that $\beta\text{-UH}_3$ retains its cubic structure up to at least 50~GPa~\cite{halevy_high_2004,kruglov_uranium_2018,guigue_synthesis_2020}, providing a broad pressure range over which both the U–U and U–H distances can be continuously reduced.

The aim of the present study is to extend the determination of the pressure dependence of $T_C$ in $\beta\text{-UH}_3$ to pressures more than an order of magnitude higher than previously achieved. We first present a method for synthesizing high-quality $\beta\text{-UH}_3$ crystals directly inside a diamond anvil cell. We then track the magnetic transition during isobaric warming using nitrogen-vacancy (NV) center magnetometry. Combined with a robust statistical analysis of the magnetic stray field, this approach enables us to determine the evolution of $T_C$ over the entire investigated pressure range, up to 20~GPa.

\section{Experimental setup and methods}

$\beta\text{-UH}_3$ is pyrophoric and must therefore be handled under an inert atmosphere~\cite{le_guyadec_pyrophoric_2010}. Once synthesized, however, the $\beta\text{-UH}_3$ phase is stable at ambient pressure and temperature, allowing the synthesis to be performed prior to the magnetic measurements.

Sample preparation was carried out in an argon-filled glove box in two steps. First, bulk uranium was mechanically cleaned to remove the surface oxide layer. Small pieces, a few micrometers in size, were then produced using a diamond file, collected, and loaded into a DAC for reaction with hydrogen under high-pressure conditions. The sample chamber was hermetically sealed inside the glove box, and the DAC was subsequently transferred to a high-pressure vessel for hydrogen loading at 140~MPa, following the procedure described in Ref.~\cite{guigue_synthesis_2020}.
X-ray diffraction (XRD) measurements were performed to monitor the chemical evolution of the U sample. The diffraction patterns revealed the spontaneous formation of $\beta\text{-UH}_3$, together with residual uranium metal, uranium oxide, and a small fraction of $\alpha\text{-UH}_3$, the latter being commonly observed as an intermediate phase during the synthesis of $\beta\text{-UH}_3$~\cite{kruglov_uranium_2018,taylor_ab_2010}. A YAG laser beam was subsequently focused onto the sample inside the DAC, heating it to approximately 1500~K for several minutes. This laser-heating treatment resulted in complete conversion to pure $\beta\text{-UH}_3$, as confirmed by the XRD pattern shown in Figure~1b.

Crystals with a typical size of approximately 10~$\mu$m were recovered by unloading the DAC inside the glove box. We selected one of those crystals and placed it, together with a temperature-independent SrB$_4$O$_7$:Sm$^{2+}$ pressure gauge~\cite{datchi_improved_1997}, in a second DAC specifically adapted to NV magnetometry~\cite{lesik_magnetic_2019}. This DAC was then was loaded with argon as the pressure-transmitting medium. After completion of the magnetic measurements, XRD analysis confirmed that the sample remained phase-pure and exhibited neither oxidation nor hydrogen loss.

\begin{figure}
\includegraphics[width = 0.48\textwidth]{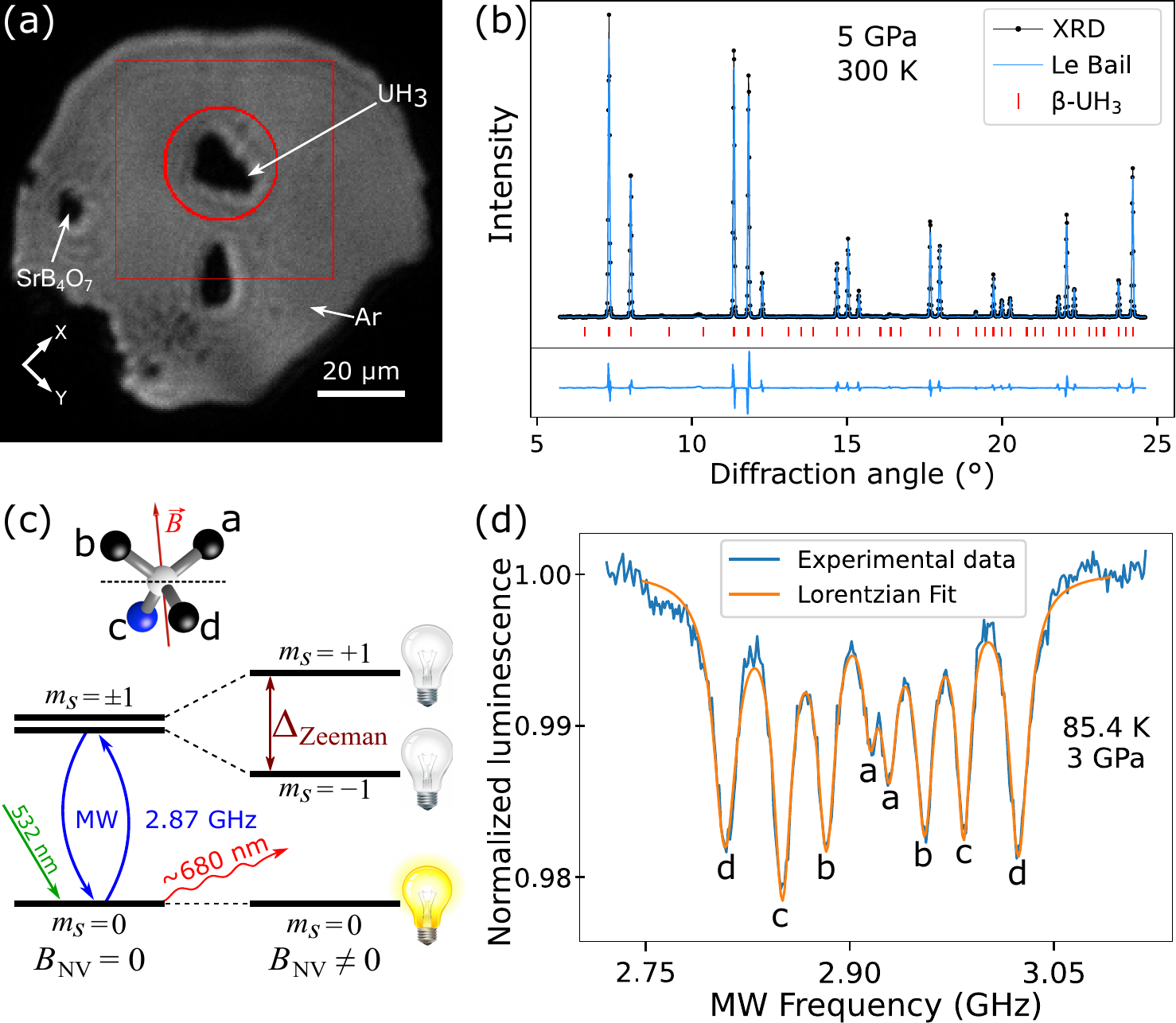}
\caption{\label{fig:Fig1} (a) White light optical image of the DAC chamber at 3~GPa. The $\beta\text{-UH}_3$ sample is highlighted, as well as the SrB$_4$O$_7$:Sm$^{2+}$ pressure gauge. (b) XRD patterns obtained after laser heating on the $\beta\text{-UH}_3$ sample. The excellent Le Bail refinement indicates the high purity of the sample. (c) Simplified energy level structure of the NV center. By combining optical and microwave excitation with optical collection, it is possible to measure the frequency splitting between the energy levels of the NV centers, which is directly linked to the magnetic field through the Zeeman effect. (d) ODMR spectrum taken near the sample at 3~GPa and 85.4~K with an applied static magnetic field of around 4~mT. A multi-lorentzian fit is performed to extract the frequency splitting values. Each pair of peaks corresponds to a crystalline orientation of the diamond lattice, indicated by letters.}
\label{fig:synthesis}
\end{figure}

To monitor the magnetic transition of $\beta\text{-UH}_3$ under pressure, we employed widefield nitrogen-vacancy (NV) center magnetometry~\cite{rondin_magnetometry_2014}, which is compatible with diamond anvil cells~\cite{lesik_magnetic_2019, hsieh_imaging_2019, yip_measuring_2019}. This technique allows for local magnetic field imaging, which avoids the background signal from the DAC and enables quantitative magnetic measurements on microscopic samples. The NV center is a point defect in diamond consisting of a substitutional nitrogen atom adjacent to a vacancy. Its spin-triplet ground state is optically initialized and read out through optically detected magnetic resonance (ODMR), while an external magnetic field lifts the degeneracy of the $\left|m_s=\pm1\right\rangle$ states through the Zeeman effect (Figure~\ref{fig:Fig1}c). In the low-field regime used here (below approximately 10~mT), the measured ODMR frequency splitting is directly proportional to the magnetic field projection along the NV axis. A shallow ensemble of implanted NV centers~\cite{lesik_maskless_2013}, with a density of approximately $10^4$~NV centers per $\mu$m$^2$, provides simultaneous measurements along the four crystallographic orientations of the diamond lattice. Combined with widefield optical imaging~\cite{scholten_widefield_2021}, this enables quantitative mapping of the magnetic stray field generated by the sample with micrometer spatial resolution~\cite{lesik_magnetic_2019,toraille2018}.

The experimental setup follows the NV-DAC configuration previously reported in Refs.~\cite{lesik_magnetic_2019, dailledouze_imaging_2025}. It consists of a (100)-oriented diamond anvil implanted with NV centers and a slitted gasket that enables microwave excitation. The NV-DAC was placed in a liquid-nitrogen cryostat, while the sample temperature was monitored using a thermocouple attached to the NV diamond anvil. The photoluminescence of the NV ensemble was imaged using a widefield microscope. A static magnetic field of approximately 4~mT was applied during the experiments.

At each pressure between 2.5 and 19.6~GPa, the sample was cooled to approximately 100~K and then continuously warmed under isobaric conditions through the ferromagnetic transition. During each warming cycle, approximately 80~ODMR maps were acquired over the temperature range of 100–180~K. The pressure was determined using either the SrB$_4$O$_7$:Sm$^{2+}$ pressure gauge~\cite{datchi_improved_1997} or the calibrated pressure dependence of the average NV ODMR frequency~\cite{hilberer_enabling_2023}. Since the pressure evolves slightly during the warming cycle, the value reported for each transition is the one obtained from the average ODMR frequency on the map acquired at the temperature closest to the measured $T_C$, with an estimated uncertainty of $0.2$~GPa. The resulting dataset therefore consisted, for each pressure point, of a sequence of magnetic stray-field maps spanning the entire magnetic transition.

\section{Detection of the Ferromagnetic Transition}
\subsection{Magnetic field reconstruction}

The vector magnetic field reconstruction, used to extract the magnetic dipole moment of $\beta\text{-UH}_3$, is illustrated in Fig.~2a, at 3.0~GPa for three temperatures. These vector magnetic field maps were obtained by doing a multi-Lorentzian fit on each pixel of a 50-$\mu$m square centered around the $\beta\text{-UH}_3$ sample (indicated in Figure~1a), yielding the stray magnetic field created by the sample magnetization. As already observed in NV magnetometry of metallic samples~\cite{lesik_magnetic_2019}, microwave screening beneath the sample can reduce the ODMR contrast in its immediate vicinity, preventing a reliable extraction of the resonance frequency splittings for the associated pixels. We chose to exclude these pixels from the reconstruction and represent in white the corresponding masked region at the center of the maps. The masked area is largest in the ferromagnetic phase, where the magnetic signal is strongest, and progressively shrinks as the sample approaches the paramagnetic state. Well below the Curie temperature ($T_C=160.1$~K at this pressure), a characteristic dipolar lobe-shaped stray-field pattern is clearly observed, demonstrating the ferromagnetic ordering of $\beta\text{-UH}_3$. As the temperature approaches the magnetic transition, the stray-field amplitude progressively decreases, reflecting the reduction of the sample magnetization. In addition, a slight rotation of the dipolar pattern is observed. This rotation results from the magnetic anisotropy weakening more rapidly than the magnetic moment close to the transition, causing the equilibrium magnetization direction to change before the complete disappearance of the magnetic order~\cite{lin_magnetic_1956}. Above the Curie temperature, the dipolar signature vanishes, indicating the transition to the paramagnetic state.

\begin{figure*}
\includegraphics[width = \textwidth]{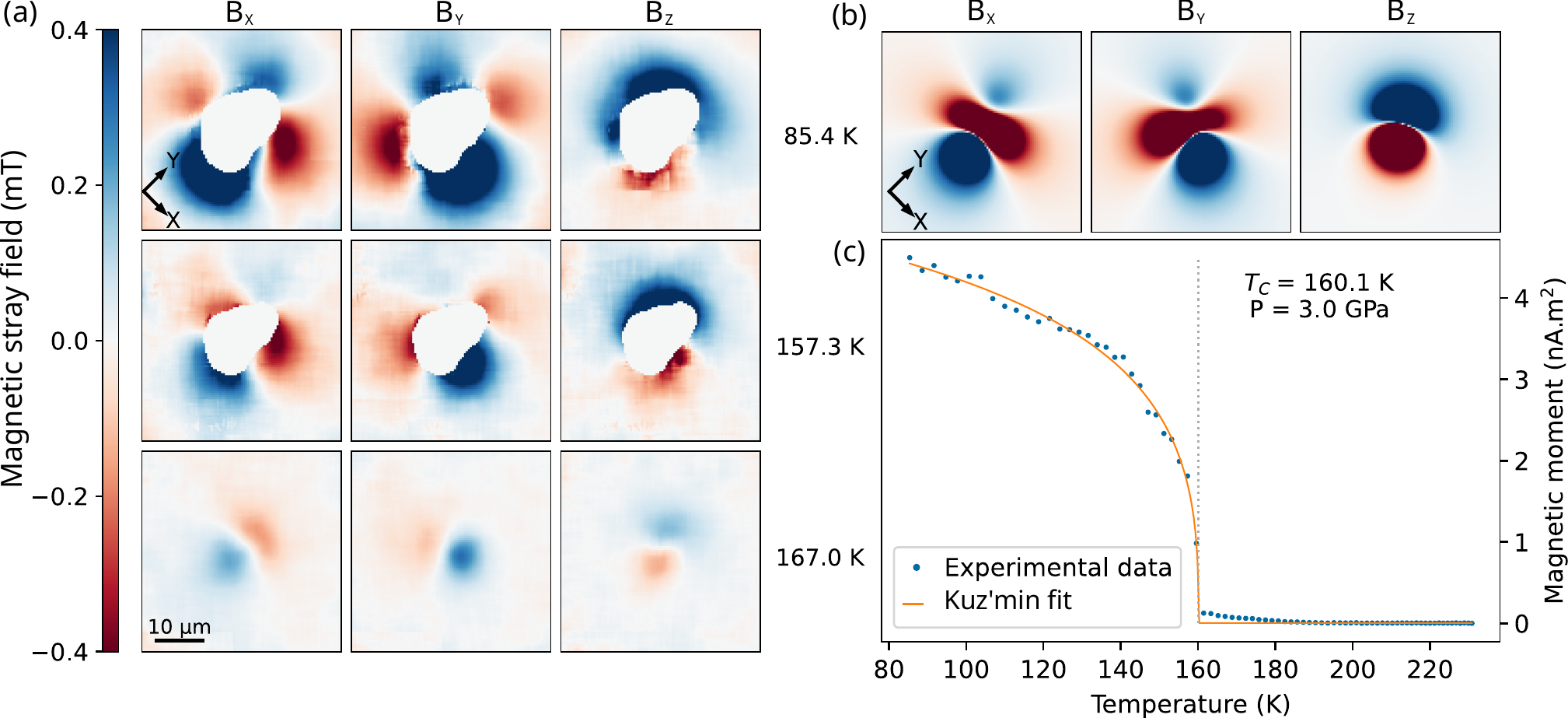}
\caption{\label{fig:Dipole} (a) Selected stray field reconstructions obtained from the experimental data taken around the sample near 3.0~GPa, for 85.4~K (top), 157.3~K (middle) and 167.0~K (bottom). The progressive transition from the ferromagnetic state at low temperature to the paramagnetic state at higher temperature is clearly visible. The white masked region corresponds to degraded ODMR spectra beneath the sample and progressively shrinks as the magnetic moment decreases. (b) Dipolar modelization of the stray field map obtained at 85.4~K, yielding a value of 4.5\,(1)~nA.m$^2$ for the magnetic moment of the sample. (c) Temperature dependence of the total magnetic moment of the sample at 3.0~GPa. A Curie temperature value $T_C=160.1\,(2)$~K can be obtained using the adjustment of equation \ref{eq:modelTc} ($s = 0.9$ and $p = 8.2$).}
\end{figure*}

The quantitative reconstruction of the vector magnetic field enables the extraction of the total magnetic moment of the sample. Assuming that the stray field $\vec{B}$ is dominated by a magnetic dipole contribution, it can be modeled using: 

\begin{equation}
    \vec{B} = \dfrac{\mu_0}{4\pi}\left( \dfrac{3(\vec{\mu}\cdot\vec{r}) \vec{r}}{|\vec{r}|^5} - \dfrac{\vec{\mu}}{|\vec{r}|^3}\right),
\end{equation}

\noindent where $\vec{\mu}$ is the magnetic moment of the sample and $\vec{r}$ is the position vector relative to the dipole center.

The result of this fitting procedure for the low temperature point (85.4~K) at 3.0~GPa is presented in Figure~\ref{fig:Dipole}b. The masked region, confined to the center of the maps, leaves unaffected the outer part of the dipolar lobes that constrain the fit. Because of the irregular sample geometry, its volume cannot be determined with sufficient accuracy to reliably estimate the magnetization value. We therefore restrict our analysis to the total magnetic moment, which is directly obtained from the magnetic field amplitude and does not depend on the sample volume. By repeating this fitting procedure for each stray field map collected during warming, we obtain the temperature dependence of the magnetic moment of the sample, shown in Figure~\ref{fig:Dipole}c. The magnetic moment progressively decreases upon approaching the magnetic transition and becomes much weaker when the sample enters the paramagnetic phase. The temperature dependence is fitted using the analytical representation proposed by Kuz'min \textit{et al.}~\cite{kuzmin_shape_2005}:
 
\begin{equation}
 \label{eq:modelTc}
M(t) = M_0\left[ 1-s \left(\frac{T}{T_C}\right)^{3/2} -(1-s)\left(\frac{T}{T_C}\right)^p\right]^{1/3},
\end{equation}
 
\noindent where $s$ and $p$ are adjustable parameters. This procedure gives a Curie temperature of $T_C = 160.1\,(2)$~K at 3.0~GPa and a saturation magnetic moment of $M_0 = 5.1\,(1)$~nA.m².

The uncertainty in the vector magnetic field reconstruction is directly related to the local signal-to-noise ratio of the ODMR spectra, since the vector components are derived from fitted resonance frequency splittings \cite{toraille2018}. Spatial variations in ODMR contrast and noise therefore produce nonuniform uncertainties across the field of view, and reliable pixel-by-pixel reconstruction becomes difficult in regions with a low signal-to-noise ratio. In addition, reconstructing the complete vector field for each ODMR map requires computationally demanding pixel-by-pixel fitting. However, determining $T_C$ only requires tracking the disappearance of the magnetic signal generated by the sample, rather than quantitatively reconstructing the three vector components of the magnetic field. This observation motivates the development of a complementary statistical analysis based on a reduced subset of the ODMR data.

\subsection{Statistical determination of the Curie Temperature}

Inspection of the stray-field maps (Fig.~\ref{fig:Dipole}a) shows that the magnetic transition can already be followed through the evolution of a single magnetic-field projection, without reconstructing the complete vector field. We therefore selected one of the four NV orientations, labeled $c$, and tracked its corresponding ODMR frequency splitting $(\Delta_c)$. This splitting measures the projection of the magnetic field onto the associated NV axis and exhibits a pronounced dipolar signature around the sample. Among the four NV orientations, orientation $c$ provided one of the largest magnetic-field projections while maintaining a stable ODMR contrast throughout the experiment, yielding the highest signal-to-noise ratio for the statistical analysis.

\begin{figure}
\includegraphics[width = 0.48\textwidth]{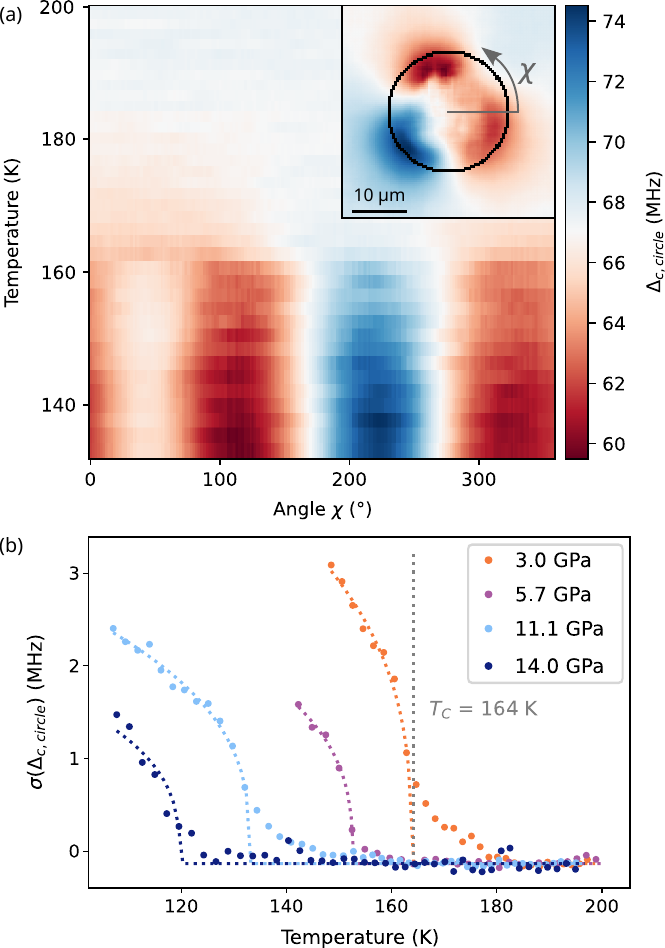}
\caption{\label{fig:Transition} (a) Evolution with temperature of $\Delta_{c,circ}$, a given frequency splitting along the $c$ direction corresponding to the NV centers orientation with the second highest magnetic projection, integrated along the angular position $\chi$ on a circle around the sample, at 3.0~GPa. The oscillations due to the presence of the magnetic dipole can be seen below the Curie temperature and disappear above it. Inset: spatial map of the corresponding frequency splitting $(\Delta_c)$ at 126.8~K. The black circle indicates the pixels of this map used for the integration. (b) Selected temperature dependences of $\sigma(\Delta_{c,circ})$, the standard deviation of the integrated frequency splitting $\Delta_{c,circ}$ at 3.0, 5.7, 11.1 and 14~GPa. A Curie temperature value $T_C=164\,(2)$~K can be obtained using the adjustment equation \ref{eq:modelTc} for the 3.0~GPa curve.}
\end{figure}

Figure~\ref{fig:Transition}a shows the spatial distribution map of $(\Delta_c)$ in the ferromagnetic phase at 126.7~K. To further reduce the dimensionality of the data subset, we select a circular contour passing through the dipolar lobes of the stray magnetic field. This contour maximizes the sensitivity to the dipolar field while avoiding the central region, where microwave screening reduces the ODMR contrast. Figure~\ref{fig:Transition}a presents the variation of the frequency splitting thus integrated along the angular position $\chi$ on this circle, denoted $\Delta_{c,circ}$ and its evolution with temperature. This representation allows a direct visual depiction of the disappearance of the magnetic dipole with temperature. At low temperatures, $\Delta_{c,circ}$ presents a strong oscillation with $\chi$, due to the presence of the four lobes of the sample magnetic dipole. The oscillation amplitude decreases as the sample approaches the Curie temperature, due to the reduction of the sample magnetic moment. At the Curie temperature, the transition from ferromagnetic to paramagnetic results in an almost complete disappearance of the oscillation, indicating the loss of ferromagnetic order. The small rotation of the stray magnetic field direction, beginning around 10~K before the transition, can also be more clearly observed on this representation as a phase shift of the oscillation pattern. For temperatures above the magnetic transition, the stray field generated by the sample vanishes and $\Delta_{c,\mathrm{circ}}$ becomes nearly independent of the angular position. The remaining frequency splitting is determined by the projection of the applied static bias field along NV orientation $c$.

To quantify the evolution of $\Delta_{c,circ}$, we consider its standard deviation $\sigma(\Delta_{c,circ})$. For a fixed orientation of the magnetic moment $\mu$, the stray field generated by the sample is linearly proportional to the magnetic moment amplitude. Since the ODMR frequency splitting is proportional to the projection of this field along the NV axis, the angular modulation of $\Delta_{c,circ}$ can be written as $\Delta_{c,circ}(\chi)-\langle\Delta_{c,circ}\rangle=\mu f(\chi)$, where $f(\chi)$ only depends on the geometry of the dipolar field. Consequently, the standard deviation $\sigma(\Delta_{c,circ})$ is directly proportional to the total magnetic moment of the sample. Figure~\ref{fig:Transition}b shows the temperature dependences of $\sigma(\Delta_{c,circ})$ for several pressures. Its evolution can be fitted using the same expression used for the dipolar reconstruction, and yields a Curie temperature $T_C = 164\,(2)$~K at 3.0~GPa. Close to the Curie temperature, the small rotation of the magnetic dipole slightly breaks the proportionality between $\sigma(\Delta_{c,circ})$ and the magnetic moment. However, because the dipole rotation remains limited, $\sigma(\Delta_{c,circ})$ still provides a reliable approximation of the magnetic moment evolution. The close agreement between the values obtained from the full dipolar reconstruction and from the statistical analysis validates the latter approach.  While dipole modeling provides quantitative information on the total magnetic moment, the statistical method enables a robust and efficient determination of $T_C$ over the entire pressure range investigated.

\begin{figure}
\includegraphics[width = 0.48\textwidth]{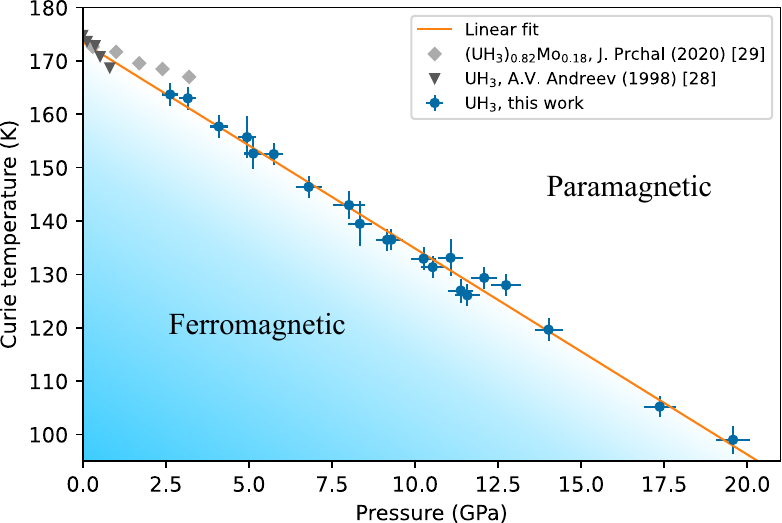}
\caption{\label{fig:CurieTemp} Pressure dependence of the Curie temperature of $\beta\text{-UH}_3$, extracted from the ODMR statistical analysis (blue cross). Error bars reflect uncertainties from the fitting procedure and temperature measurements. The solid line is a linear fit yielding a decrease $\text{d}T_C/\text{d}P = -3.86\,(8)$~K/GPa, indicating a progressive weakening of ferromagnetic properties under compression. Extrapolation to ambient pressure gives a value of $T_C = 173.5\,(9) \text{~K}$, consistent with previous studies~\cite{gruen_magnetic_1955, troc_discovery_1995, andreev_magnetic_1998, tereshina-chitrova_synthesis_2023}. The red and yellow crosses correspond to previous measurements under pressure on pure $\beta\text{-UH}_3$~\cite{andreev_magnetic_1998}, and $\beta\text{-UH}_3$ stabilized by alloying it with molybdenum~\cite{prchal_pressure_2020}.}
\end{figure}

\subsection{Pressure Evolution of Curie Temperature}

For each pressure point between 2.5~GPa and 19.6~GPa, we applied the statistical analysis procedure and obtained the Curie temperature values shown in Figure~\ref{fig:CurieTemp}. The Curie temperature decreases linearly with pressure over the investigated range. A linear fit yields a slope of $\text{d}T_C/\text{d}P = -3.86\,(8) \text{~K/GPa}$. Extrapolation of this fit to ambient pressure gives $T_C=173.5(9)$ K, in good agreement with previous measurements~\cite{gruen_magnetic_1955, troc_discovery_1995, andreev_magnetic_1998, tereshina-chitrova_synthesis_2023}. This agreement supports the accuracy of the NV magnetometry measurements under pressure.

Curie-temperature measurements could not be extended beyond 20~GPa, where $T_C$ approaches 100~K: because the statistical analysis requires data acquired at least 10~K below the Curie temperature, the minimum sample temperature of approximately 90~K attainable with our liquid nitrogen cryostat prevented reliable measurements at higher pressures.

The pressure dependence measured here is comparable to that previously reported for pure $\beta\text{-UH}_3$ up to 0.8~GPa~\cite{andreev_magnetic_1998}. However, the pronounced nonlinearity observed in that study is not reproduced in the present measurements, for which $T_C(P)$ remains linear over the investigated pressure range, up to about 20~GPa. The magnitude of the slope measured for pure $\beta\text{-UH}_3$ is approximately twice that reported for Mo-alloyed$\text{-UH}_3$~\cite{prchal_pressure_2020}. This stronger pressure dependence is consistent with a more itinerant character of the uranium 5f electrons in pure $\beta\text{-UH}_3$. Hydrostatic compression increases the 5f bandwidth and the degree of orbital hybridization, thereby destabilizing the ferromagnetic state. Conversely, the weaker pressure dependence observed in Mo-alloyed$\text{-UH}_3$ is consistent with the more localized 5f-electron character proposed for this compound.

A quantum critical point may emerge when a continuous phase transition, such as the paramagnetic-to-ferromagnetic transition, is driven to zero temperature. The observed linear decrease of $T_C$ with pressure suggests that the ferromagnetic transition in $\beta\text{-UH}_3$ could be suppressed to $0$~K at approximately 44~GPa, potentially providing access to quantum critical behavior in this compound. Importantly, X-ray diffraction studies have shown that the $\beta\text{-UH}_3$ phase remains stable over this pressure range~\cite{halevy_high_2004,kruglov_uranium_2018,guigue_synthesis_2020}, indicating that this quantum-critical regime may be experimentally accessible.

The decrease of $T_C$ with pressure raises the possibility of a pressure-induced quantum critical point in $\beta\text{-UH}_3$. Extrapolation of the measured $T_C(P)$ dependence suggests that ferromagnetic order could vanish at a critical pressure of approximately 44~GPa. This estimate nevertheless assumes that the linear pressure dependence persists beyond the investigated range, that the ferromagnetic transition remains continuous, and that no intervening structural or magnetic phase emerges. Importantly, high-pressure X-ray diffraction measurements have shown that the $\beta\text{-UH}_3$ crystal structure remains stable up to at least 50~GPa~\cite{halevy_high_2004,kruglov_uranium_2018,guigue_synthesis_2020}, suggesting that the extrapolated critical pressure range is structurally accessible. NV-based magnetic measurements performed at higher pressures and lower temperatures could determine whether ferromagnetism is continuously suppressed to zero temperature and whether quantum critical behavior actually emerges.

\section{Conclusion and outlook}

We have extended previous measurements of the magnetic phase diagram of $\beta\text{-UH}_3$ and demonstrated that its Curie temperature decreases linearly with pressure. The persistence of ferromagnetism throughout the investigated pressure range highlights the robustness of magnetic ordering in $\beta\text{-UH}_3$ and provides new constraints on the interplay between U--U interactions, U--H hybridization, and 5f-electron localization in uranium hydrides. Extending these measurements to higher pressures would provide further insight into the evolution of ferromagnetism in $\beta\text{-UH}_3$. In the present study, the accessible temperature--pressure range was primarily limited by the use of a nitrogen-cooled cryostat. NV magnetometry itself can operate without significant degradation of the magnetic signal at pressures well beyond those explored here, including the projected pressure of approximately 44~GPa at which ferromagnetism may ultimately be suppressed~\cite{hilberer_enabling_2023}.

An important open question concerns the pressure dependence of the uranium magnetic moment. In principle, this quantity could be inferred from NV measurements of the sample's magnetic stray field and related quantitatively to its magnetization, provided that the sample geometry is sufficiently well controlled. A simple and well-defined geometry, ideally close to spherical, would substantially reduce uncertainties associated with demagnetization effects and facilitate a quantitative determination of the magnetic moment. Alternatively, element-specific information on the uranium 5f magnetic moment can be obtained using X-ray magnetic circular dichroism (XMCD), which has already been applied to a broad range of uranium compounds at ambient pressure ~\cite{wilhelm_magnetism_2018,tereshina-chitrova2026}. Extending this technique to high-pressure measurements in a DAC remains technically challenging because of the microscopic sample volume and the restricted X-ray access through the anvils and gasket, particularly when low temperatures and high magnetic fields must be combined.

Beyond advancing the understanding of uranium hydride magnetism, these results further demonstrate the capability of NV magnetometry to quantitatively probe magnetic phase transitions under high pressure. In particular, the statistical analysis developed here provides a complementary and efficient approach for tracking magnetic transitions. More broadly, this methodology could be adapted to other pressure-induced phase transitions, including superconducting transitions detected through the onset of the Meissner response.

\section{Acknowledgments}

We acknowledge the European Synchrotron Radiation Facility (ESRF) for provision of synchrotron radiation facilities, and we thank Gaston Garbarino for his help, assistance and support in acquiring X-ray diffraction data using the beamline ID15B (in-house research).
We thank Martin Schmidt for performing the NV FIB implantation.
This work has been partially funded by the European Research Council with the ERC Advanced Grant “QPRESSE” (No. 101142682). J.-F.~Roch acknowledges support from the Institut Universitaire de France.

\bibliography{Biblio}

\end{document}